# Stimulated emission in erbium doped silicon rich nitride waveguides

Debo Olaosebikan[1], Selcuk Yerci[2], Alexander Gondarenko[1], Kyle Preston[1], Rui Li[2], Luca Dal Negro[2] and Michal Lipson[1*]

[1]*Department of Electrical and Computer Engineering, Cornell University, Ithaca, NY, 14853, USA*
[2]*Department of Electrical Engineering, Boston University, Boston, MA, 02215, USA*
[*]*ml292@cornell.edu*

**Abstract:** Stimulated emission of sensitized Erbium atoms is reported in silicon-rich silicon nitride waveguides. Visible pump and infrared probe measurements are carried out in waveguides fabricated from erbium-doped silicon rich silicon nitride. A decrease in the photoinduced absorption of the probe in the wavelength range of the erbium emission is observed and is attributed to stimulated emission from erbium atoms excited indirectly via the silicon excess in an $SiN_x$ matrix. A near 50% decrease in absorption is measured, corresponding to an 8% fraction of inverted erbium atoms. This fraction is an order of magnitude higher than that obtained in counterpart oxide systems possessing observable nanocrystals. Our results indicate that population inversion and gain at 1.54μm might be possible by optimizing the silicon excess in the $SiN_x$ matrix.

## 1. Introduction

The quest for CMOS-compatible light sources and amplifiers continues to drive interest and research in the emergent field of silicon photonics [1-5]. Potential platforms must be electrically pumped and should preferably emit at the technologically relevant, 1550nm telecommunications wavelength.

In this regard, erbium-doped silicon-rich silicon nitride (Er: SRN) is a promising material candidate [6]. Localized trap states as well as small nanometer-sized silicon nanocrystals in a silicon nitride host act as highly efficient sensitizers for surrounding erbium ions [7-9]. The excess silicon permits electrical access [10, 11] while the erbium ions serve as emitters at 1550 nm. Sensitized stimulated emission, erbium population inversion, and ultimately gain must be achieved in an integrated waveguide, towards the end of achieving lasing with Er: SRN as the active material.

Inverting a significant fraction of erbium ions via nanocrystals has previously been problematic with erbium inversion ratios of ~ 1% being the demonstrated state of the art [12]. Efficient photoluminescence [7-9], electroluminescence [10, 11] and even optical gain under intense pulsed conditions [13] have been demonstrated in silicon nanocrystal based systems without erbium. However, gain under continuous wave excitation has not been demonstrated the emission peak of the nanocrystals (~750nm) is far from the telecom band. In addition, while sensitized stimulated emission from erbium in silicon rich silicon oxide was reported in [14], photoinduced absorption was reported in [15]. Thus, engineering the optimal silicon rich matrix for obtaining sensitized stimulated emission and gain remains an important task to be resolved consistently before lasing can be achieved [16].

Here we present evidence of stimulated emission from indirectly sensitized erbium ions in a silicon rich nitride host matrix, and show a 50% suppression of the expected carrier related losses. This corresponds to an 8% inversion fraction of erbium. The magnitude of the suppression, and the order of magnitude increase in the inversion fraction as compared to previous work, suggests that the deleterious effects of generated carriers and the challenges of inverting erbium might not be as severe in the nitride matrix as in its oxide counterpart due to a faster excitation of erbium [7-9] and that the optimization of silicon content and pump conditions could lead to net gain.

## 2. Fabrication and Passive Testing of Er: SRN Waveguides and Resonators

The waveguides were fabricated from 400nm thick Er: SRN films. The films were deposited by reactive sputtering as detailed elsewhere [9, 10]. The sputtered film has a relative Si atomic concentration of 47% as compared to 43% in stoichiometric silicon nitride indicating a 4% silicon excess [7]. The erbium concentration has been determined from Rutherford back scattering measurements to be $\sim 4.6 \times 10^{20} cm^{-3}$ [9]. To verify the optical activity of the erbium and the sensitizing matrix, erbium photoluminescence was obtained from the films by pumping at 457nm, away from the erbium absorption line at 488nm (Fig. 1(a)). The Er: SRN was subsequently patterned via electron beam lithography, and reactive ion etching. Fig. 1(b) shows a cross section of the fabricated devices.

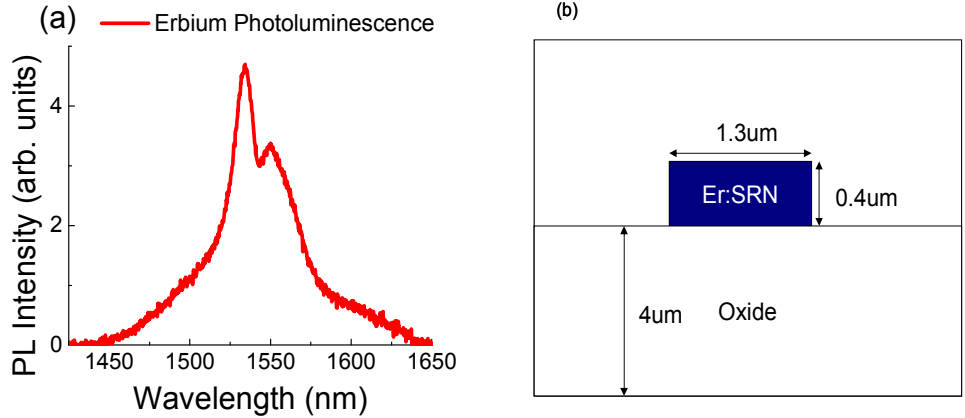

Fig. 1. (a): Sensitized erbium emission obtained by pumping at 457nm which is away from the erbium resonance absorption at 488nm. (b): Cross section of the fabricated Er:SRN waveguides with refractive index n=2.23 cladded in oxide and on top of a 4um oxide box

In these devices, scattering limited waveguide losses of ~20dB/cm are achieved. The losses were determined by measuring the transmission through near identical waveguides of different lengths. While ultimately the losses must be brought down for net amplification, the system is already good enough to measure pump-induced losses via changes in waveguide transmission.

3. **Pump-Probe Measurements of Er: SRN waveguides**

In a pump-probe setup, we extract the dependence of the photoinduced losses on the probe wavelength. A continuous wave at 470nm is used to pump the Er: SRN waveguides. The pump beam is focused with a cylindrical lens into a line directly on top of the waveguide being measured, at an estimated incident light intensity of $\sim 50 mW/cm^2$. The probe (the output of a tunable laser of picometer precision) is coupled into the waveguides and scanned over a 1.5 μm to 1.6 μm range.
   In the presence of the pump, we observe a suppression of probe transmission for all wavelengths due to photogenerated carriers in the SRN matrix (Fig. 2).

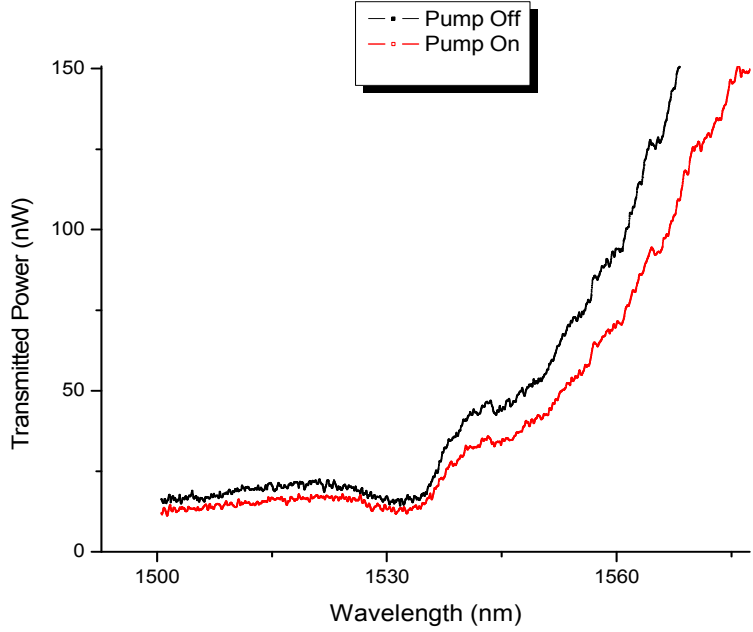

Fig. 2. The wavelength dependence of the transmission through a 7.6 mm long waveguide is depicted in pump-on and pump-off conditions.

The pump induced absorption, $\Delta\alpha$, is determined from the transmission in the pump-on and pump-off scenarios as:

$$\Delta\alpha = -\frac{1}{L}\ln\left(\frac{P_{on}}{P_{off}}\right), \quad (1)$$

$P_{on}$ and $P_{off}$ represent the magnitude of the power transmitted through the waveguide with the pump on and off while $L$ denotes the length of the waveguide. Here, a positive $\Delta\alpha$ represents an increase in absorption.

We find evidence for stimulated emission from erbium in the spectral dependence of the pump-induced absorption. When this photoinduced absorption, $\Delta\alpha$, is plotted as a function of wavelength ( Fig. 3.), there is a clear dip around the 1.54 μm erbium emission peak against an overall background loss. Such a feature is absent in similar waveguides fabricated from materials that were not co-sputtered with erbium (Fig. 3). These control waveguides were non-resonantly pumped at 455 nm where there is a higher sensitization cross section [7-9], and had 20% greater confinement of the mode for increased sensitivity to changes in waveguide absorption. Even in these favorable conditions, the suppression was not seen. Thus, we conclude that the dip is related to the presence of erbium.

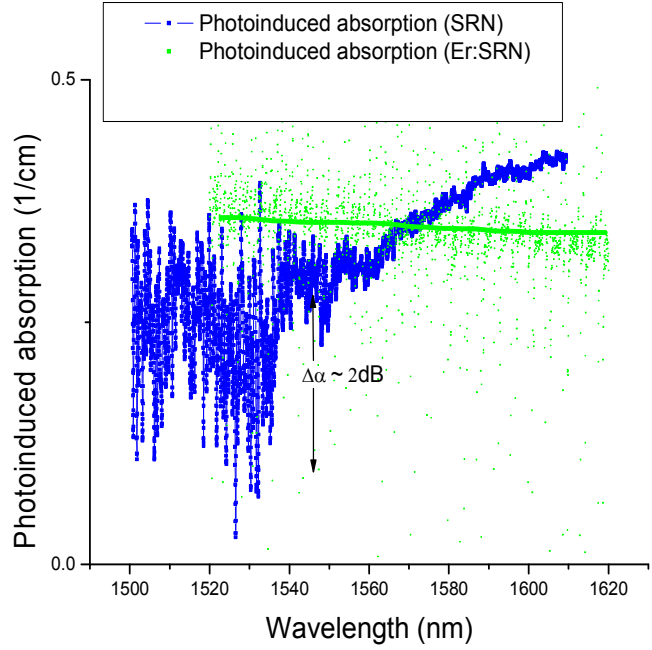

Fig. 3 The photoinduced absorption in silicon rich nitride waveguides with and without erbium ions. No dip is observed in SRN waveguides.

We note that the transmitted probe is measured directly by an InGaAs photo detector with a flat response in the wavelength regime studied. Thus any spontaneous emission generated by the pump is detected as an overall bias in the probe signal at all wavelengths. The probe beam however, being intense and coherent can stimulate transitions in the erbium population. As the probe wavelength is varied, these stimulated transitions will follow the wavelength dependence of the erbium emission cross section. This picture is confirmed in our measurement and is manifested in the similarity between the dip in the absorption and the

We deduce an 8% inverted erbium fraction from the magnitude of the absorption dip. Considering that the origin of the dip in absorption is erbium stimulated emission, the inverted erbium population is estimated as:

$$\left|\Delta\alpha_{dip}\right| = \Gamma \sigma_e N_2 \quad (2)$$

The inversion fraction, $f$, is thus given by:

$$f = \frac{N_2}{N_0} \quad (3)$$

where $\left|\Delta\alpha_{dip}\right|$ is the magnitude of the dip in the absorption, $\Gamma$ is the modal confinement factor [17], $\sigma_e$ is the erbium emission cross section at 1.54 μm, $N_2$ is the excited erbium population density and $N_0$ is the total erbium density. Using the appropriate material parameters and waveguide dimensions, we numerically obtain a value of $\Gamma = 0.67$ for the fundamental TM mode used in the experiment.

To determine the inversion fraction, we use the experimentally measured values of $|\Delta\alpha_{dip}| = 0.14\ cm^{-1}$ and $N_0 = 4.6 \times 10^{20}\ cm^{-3}$. Given the small index difference between silicon nitride and silica, and the fact that the emission cross section of erbium varies little across different oxide systems; we use a value of $\sigma_e = 6 \times 10^{-21} cm^2$, which is close to the highest measured value for erbium in silica [18]. By choosing a high value for $\sigma_e$, we obtain a lower inversion fraction and thus a more conservative estimate. Using equations (2) and (3) along with the appropriate values, we get $f \approx 0.08$ corresponding to an 8% inversion ratio.

The inversion ratio obtained is a lower limit to the true ratio since not all erbium ions are optically active and the pump intensities accessible in the experiment were not enough to saturate the absorption dip. Both factors imply that the inverted fraction measured in this experiment is not a fundamental limit and inversion might yet be possible. However, the 8% inversion ratio is an order of magnitude greater than that measured in silica host systems [12]. The method used in this work, examining the wavelength dependence of the photoinduced absorption to extracting the excited erbium fraction, can be extended to oxide systems as well and is a fairly simple experimental tool for probing the fraction of inverted atoms in real devices made from these complicated material systems.

The higher amount of erbium inversion measured in the Er: SRN system could be due to the faster erbium sensitization in the nitride. Transfer times from the SRN matrix to the erbium ions are on the order of nanoseconds as compared to the microsecond times in oxide systems [7-9]. Further studies of the temporal dynamics of erbium excitation and carrier generation in waveguide based structures are needed to clarify the limits to the inversion ratio and whether or not gain is possible.

## 5. Conclusions

The results presented suggest less silicon rich samples than the one studied here might lead to a smaller overall loss background that allows the dip seen in this experiment cross over into the region of negative absorption (gain). Decreasing the silicon excess in the material must be done with caution since the silicon network is needed for electrical access and excitation of at least 50% of the erbium. Thus, a careful optimization over a range of silicon content, and pump powers must be carried out to investigate the possibility of gain in these waveguides under continuous wave conditions.

**Acknowledgements**
The authors acknowledge C. Manolatou for the use of her FD code and thank Long Chen, Gustavo Wiederhecker, and Prof. Farhan Rana for productive discussions. This work was funded in part by the U.S. Air Force MURI program on "Electrically Pumped Silicon-Based Lasers for Chip-Scale Nanophotonic Systems" supervised by Dr. Gernot Pomrenke. This work was performed in part at the Cornell Nanoscale Facility, a member of the National Nanotechnology Infrastructure Network, which is supported by the National Science Foundation.